\RequirePackage{ifpdf}
\ifpdf 
\documentclass[pdftex]{sigma}
\else
\documentclass{sigma}
\fi

\usepackage{dsfont}
\numberwithin{equation}{section}

\def\ri{{\rm{i}}}
\def\tr{\mathop{\hbox{\rm tr}}\nolimits}
\def\alg{{\mathfrak g}}
\def\su2{{\mathfrak {su}}(2)}
\def\e3{{\mathfrak {e}}(3)}
\def\half{\frac{1}{2}}

\def\be{\beta}
\def\al{\alpha}
\def\ep{\varepsilon}
\def\l{\lambda}
\def\la{\lambda}

\def\sig{\sigma}

\newcommand{\CH}{{\mathcal H}}
\newcommand{\CB}{{\mathcal B}}

\newcommand{\CL}{{\mathcal L}}

\newcommand{\CG}{{\mathcal G}}
\newcommand{\CA}{{\mathcal A}}
\newcommand{\CM}{{\mathcal M}}

\begin{document}

\allowdisplaybreaks

\renewcommand{\PaperNumber}{058}

\renewcommand{\thefootnote}{$\star$}

\FirstPageHeading

\ShortArticleName{From $\su2$ Gaudin Models to Integrable Tops}

\ArticleName{From $\boldsymbol{\su2}$ Gaudin Models to Integrable
Tops\footnote{This paper is a contribution to the Vadim Kuznetsov
Memorial Issue `Integrable Systems and Related Topics'. The full
collection is available at
\href{http://www.emis.de/journals/SIGMA/kuznetsov.html}{http://www.emis.de/journals/SIGMA/kuznetsov.html}}}

\Author{Matteo PETRERA~$^\dag$ and Orlando RAGNISCO~$^\ddag$}

\AuthorNameForHeading{M. Petrera and O. Ragnisco}

\Address{$^\dag$~Zentrum Mathematik, Technische Universit\"at
M\"unchen,\\
$\phantom{^\dag}$~Boltzmannstr. 3, D-85747 Garching bei M\"unchen, Germany}
\EmailD{\href{mailto:petrera@ma.tum.de}{petrera@ma.tum.de}}

\Address{$^\ddag$~Dipartimento di Fisica E. Amaldi,
Universit\`a degli Studi Roma Tre and Sezione INFN,\\
$\phantom{^\ddag}$~Roma Tre, Via della Vasca Navale 84, 00146
Roma, Italy}
\EmailD{\href{mailto:ragnisco@fis.uniroma3.it}{ragnisco@fis.uniroma3.it}}

\ArticleDates{Received March 13, 2006; Published online April 20,
2007}

\Abstract{In the present paper we derive two well-known integrable
cases of rigid body dynamics (the Lagrange top and the Clebsch
system) performing an algebraic contraction on the two-body Lax
matrices governing the (classical) $\su2$ Gaudin models. The
procedure preserves the linear $r$-matrix formulation of the
ancestor models. We give the Lax representation of the resulting
integrable systems in terms of $\su2$ Lax matrices with rational
and elliptic dependencies on the spectral parameter. We f\/inally
give some results about the many-body extensions of the
constructed systems.}


\Classification{70E17; 70E40; 37J35}

\begin{flushright}
{\it Dedicated to the memory of Vadim B. Kuznetsov (1963--2005)}
\end{flushright}

\renewcommand{\thefootnote}{\arabic{footnote}}
\setcounter{footnote}{0}

\section{Introduction}

The Gaudin models were introduced in 1976 by M. Gaudin \cite{G1}
and attracted considerable interest among theoretical and
mathematical physicists, playing a distinguished role in the realm
of integrable systems. Their peculiar properties, holding both at
the classical and at the quantum level, are deeply connected with
the long-range nature of the interaction described by its
commuting Hamiltonians, which in fact yields a typical ``mean
f\/ield'' dynamics.

Indeed the Gaudin models describe completely integrable classical
and quantum long-range spin chains. The original Gaudin model was
formulated as a quantum spin model related to the Lie algebra
$\mathfrak{su}(2)$ \cite{G1}. Later it was realized that such
models can be associated with any semi-simple complex Lie algebra
$\alg$ \cite{G2,J} and a solution of the corresponding classical
Yang--Baxter equation \cite{BD,S0}. An important feature of Gaudin
models is that they can be formulated in the framework of the
$r$-matrix approach. In particular, they admit a linear $r$-matrix
structure that characterizes both the classical and the quantum
models, and holds whatever be the dependence (rational (XXX),
trigonometric (XXZ), elliptic (XYZ)) on the spectral parameter. In
this context, it is possible to see Gaudin models as appropriate
``semiclassical" limits of the integrable Heisenberg magnets
\cite{ST}, which admit a quadratic $r$-matrix structure.

In the 80's, the rational Gaudin model was studied by Sklyanin
\cite{S1} and Jur\v{c}o \cite{J} from the point of view of the
quantum inverse scattering method. Precisely, Sklyanin studied the
$\su2$ rational Gaudin models, diagonalizing the commuting
Hamiltonians by means of separation of variables and stressing the
connection between his procedure and the functional Bethe Ansatz.
On the other hand, the algebraic structure encoded in the linear
$r$-matrix algebra allowed Jur\v{c}o to use the algebraic Bethe
Ansatz to simultaneously diagonalize the set of commuting
Hamiltonians in all cases when $\alg$ is a semi-simple Lie
algebra. We have to mention here also the the work of Reyman and
Semenov-Tian-Shansky \cite{RSTS}. Classical Hamiltonian systems
associated with Lax matrices of the Gaudin-type were
 studied by them in the context of a general group-theoretic approach.

Vadim Kuznetsov, to whom this work is dedicated, widely studied
Gaudin models, especially from the point of view of their
separability properties \cite{KKM,KKM2,K} and of their integrable
discretizations through B\"acklund transformations \cite{HKR,KPR}.
In \cite{KPR} we collaborated with him  showing that the Lagrange
top can be obtained through an algebraic contraction procedure
performed on the two-body $\su2$ rational Gaudin model. Such a
derivation of the Lagrange system preserves the linear $r$-matrix
algebra of the ancestor model, and it has been used as a tool to
construct an integrable discretization starting from a known one
for the rational $\su2$ Gaudin model \cite{HKR}.

The purpose of the present paper is twofold: on one hand we recall
the procedure we used in \cite{KPR} to obtain the Lagrange top
from the two-body $\su2$ rational Gaudin model; on the other hand
we show how the same technique can be used to derive a special
case of the Clebsch system (i.e. the motion of a free rigid body
in an ideal incompressible f\/luid) starting from the elliptic
$\su2$ Gaudin model. In the last Section we show how to construct
many-body extensions starting from the obtained Lax matrices
governing the Lagrange top and the Clebsch system.

\section[A short review of $su(2)$ Gaudin models]{A short review of $\boldsymbol{\su2}$ Gaudin models}

The aim of this Section is to give a terse survey of the main
features of $\su2$ Gaudin models. In particular we shall describe
them in terms of their (linear) $r$-matrix formulation, providing
their Lax matrices and $r$-matrices. For further details we remand
at the references \cite{G1,G2,HKR,J,KKM,MPR,P,RSTS,S1,S3,ST}.

Let us choose the following basis of the linear space $\su2$:
\[
\sigma_1 \doteq  \half \left(\begin{array}{cc}
0 & -{\rm{i}}  \\
-{\rm{i}} & 0
\end{array}\right), \qquad
\sigma_2 \doteq  \half \left(\begin{array}{cc}
0 & -1  \\
1 & 0
\end{array}\right), \qquad
\sigma_3 \doteq \half \left(\begin{array}{cc}
-{\rm{i}} & 0  \\
0 & {\rm{i}}
\end{array}\right).
\]
We recall that the correspondence
\[
\mathbb{R}^3 \ni {\bf{a}} \doteq (a^1,a^2,a^3) \;
\longleftrightarrow \; {\bf{a}} \doteq \half
\left(\begin{array}{cc}
-{\rm{i}} \,a^{3} &  -{\rm{i}} \,a^{1}-a^2 \\
-{\rm{i}} \,a^{1}+ a^2 & {\rm{i}} \, a^{3}
\end{array}\right) \, \in \mathfrak{su}(2),
\]
is an isomorphism between $(\mathfrak{su}(2),[\,\cdot, \cdot\,])$
and the Lie algebra $(\mathbb{R}^3, \times)$, where $\times$
stands for the vector product. This allows us to identify
$\mathbb{R}^3$ vectors and $\su2$ matrices. We supply
$\mathfrak{su}(2)$ with the scalar product $\langle \, \cdot,
\cdot \, \rangle$ induced from $\mathbb{R}^3$, namely $ \langle \,
{\bf{a}}, {\bf{b}}  \, \rangle = -2 \, {\rm{tr}} \, ({\bf{a}}\,
{\bf{b}}) = 2 \, {\rm{tr}} \, ( {\bf{b}} \, {\bf{a}}^\dag), \;
\forall \,  {\bf{a}}, {\bf{b}}  \in \su2. $ This scalar product
allows us to identify the dual space $\mathfrak{su}^*(2)$ with
$\su2$, so that the coadjoint action of the algebra becomes the
usual Lie bracket with minus.

The Lie--Poisson algebra of the $N$-body $\su2$ Gaudin models is
given by (minus) $\oplus^N \mathfrak{su}^*(2)$. We will denote by
$\{y^\al_i \}_{\al=1}^{3}$, $1 \leq i \leq N$, the set of the
(time-dependent) coordinate functions relative to the $i$-th copy
of $\su2$.  Consequently, the Lie--Poisson brackets on $\oplus^N
\mathfrak{su}^*(2)$ read
\begin{gather}
\big\{ y^\al_i, y^\be_j \big\}= -  \delta_{i,j} \, \epsilon_{\al
\be \gamma} \, {y}^\gamma_{i}, \label{LPgdd}
\end{gather}
with $1\leq i,j \leq N$. Here  $\epsilon_{\al \be \gamma}$ is the
skew-symmetric tensor with $\epsilon_{1 2 3}=1$. The brackets
(\ref{LPgdd}) are degenerate: they possess the $N$ Casimir
functions{\samepage
\begin{gather}
C_{i}  \doteq \half \langle \, {\bf{y}}_i,{\bf{y}}_i \, \rangle,
\qquad 1 \leq i \leq N, \label{poi}
\end{gather}
that provide a trivial dynamics.}

The $\mathfrak{su}(2)$ rational, trigonometric and elliptic Gaudin
models are governed respectively by the following Lax matrices
def\/ined on the loop algebra $\su2 [\,\l,\l^{-1}]$:
\begin{gather}
\CL_{\CG}^r(\l)  \doteq \sig_\al\, p^\al + \sum_{i=1}^{N}
\frac{\sig_\al\, y_i^\al }{\l - \l_i}=
{\bf{p}} + \sum_{i=1}^{N} \frac{{\bf{y}}_i }{\l - \l_i} , \label{lgaure} \\
\CL_{\CG}^t(\l)  \doteq \sum_{i=1}^{N} \frac{1}{\sin (\l - \l_i)}
\left[\, \sig_1\, y_i^1 + \sig_2\, y_i^2+
\cos(\l -\l_i) \, \sig_3\, y_i^3 \, \right] , \label{lgauretrig} \\
\CL_{\CG}^e(\l)  \doteq \sum_{i=1}^{N} \frac{1}{{\rm sn} (\l -
\l_i)} \left[\, {\rm dn} (\l -\l_i) \, \sig_1\, y_i^1+ \sig_2\,
y_i^2 + {\rm cn} (\l -\l_i) \, \sig_3\, y_i^3  \, \right],
\label{lgaurell}
\end{gather}
where the $\lambda_i$'s, with $\lambda_i \neq \lambda_k$, $1 \leq
i,k \leq N$, are complex parameters of the model. We remark that
in equation (\ref{lgaurell}) $\rm{cn}(\l),\rm{dn}(\l),\rm{sn}(\l)$
are the elliptic Jacobi functions of modulus $k$. In
equation~(\ref{lgaure}) ${\bf{p}}$ is a constant vector in
$\mathbb{R}^3$. Its presence is necessary in the rational case in
order to get a~suf\/f\/icient number of functionally independent
integrals of motion.

It is well-known that the Lax matrices (\ref{lgaure}),
(\ref{lgauretrig}) and (\ref{lgaurell}) describe completely
integrable systems on the Lie--Poisson manifold associated with
$\oplus^N \mathfrak{su}^*(2)$.  In particular they admit a linear
$r$-matrix formulation, which ensures that all the spectral
invariants of  $ \CL_{\CG}^r(\l)$, $\CL_{\CG}^t(\l)$,
$\CL_{\CG}^e(\l)$ form a family of involutive functions. Let us
give the following result.

\begin{proposition} The Lax matrices $ \CL_{\CG}^r(\l)$, $\CL_{\CG}^t(\l)$, $\CL_{\CG}^e(\l)$
given in equations \eqref{lgaure}, \eqref{lgauretrig} and
\eqref{lgaurell} satisfy the linear $r$-matrix algebra
\begin{gather}
\big\{ \CL_\CG^{r,t,e}(\la) \otimes \mathds{1}, \mathds{1} \otimes
\CL_\CG^{r,t,e}(\mu) \big\}+\big[\,r_{r,t,e}(\la-\mu),
\CL_\CG^{r,t,e}(\la) \otimes \mathds{1}+\mathds{1} \otimes
\CL_\CG^{r,t,e}(\mu)\, \big]=0, \label{pb4}
\end{gather}
for all $\l, \mu \in \mathbb{C}$, with
\begin{gather}
r_{r,t,e}(\l) \doteq - f^\al_{r,t,e}(\l) \, \sigma_\al \otimes
\sigma_\al, \label{pb43ui}
\end{gather}
and
\begin{gather*}
f_r^\al(\l) \doteq \frac{1}{\l} \qquad \forall \, \al=1,2,3 , \nonumber \\
(f_t^1(\l),f_t^2(\l),f_t^3(\l)) \doteq \left(\frac{1}{\sin (\l) }, \frac{1}{\sin (\l) }, \cot (\l)\right),  \nonumber \\
(f_e^1(\l),f_e^2(\l),f_e^3(\l)) \doteq
\left(\frac{\rm{dn}(\l)}{\rm{sn}(\l)},
  \frac{1}{\rm{sn}(\l)}, \frac{\rm{cn}(\l)}{\rm{sn}(\l)} \right).  \nonumber
\end{gather*}
In equation \eqref{pb4} $\mathds{1}$ denotes the $2\times2$
identity matrix and $\otimes$ stands for the tensor product in
$\mathbb{C}^2 \otimes \mathbb{C}^2$.
\end{proposition}

In the rational case the $r$-matrix is equivalent to $r_r(\l)=-\Pi
/ (2 \, \l)$, where $\Pi$ is the permutation operator in
$\mathbb{C}^2 \otimes \mathbb{C}^2$.

The complete set of integrals of the $\mathfrak{su}(2)$ rational,
trigonometric and elliptic Gaudin models can be constructed
computing the residues in $\l=\l_i$ of the characteristic curve
$\det( \CL_{\CG}^{r,t,e}(\l) -\mu \, \mathds{1})=0$ (or
equivalently $\mu^2 = -(1/2) \tr[\,(\CL_{\CG}^{r,t,e}(\l)  )^2]$).
The following results hold.

\begin{proposition} The hyperelliptic curve $\det( \CL_{\CG}^{r}(\l) -\mu \, \mathds{1})=0$,
$\l,\mu \in \mathbb{C}$, with $\CL_{\CG}^r(\l)$ given in equation
\eqref{lgaure}, provides a  set of $2N$ independent involutive
integrals of motion given by
\begin{gather}
H_{i}^r \doteq \langle \,  {\bf{p}} , {\bf{y}}_i \, \rangle +
\sum_{\stackrel{\scriptstyle{j=1}} {j \neq i}}^N \frac{\langle \,
{\bf{y}}_i,{\bf{y}}_j \, \rangle }{\l_i - \l_j},  \qquad
\sum_{i=1}^N  H_i^r = \sum_{i=1}^N \langle \,  {\bf{p}} ,
{\bf{y}}_i \, \rangle, \label{35}
\end{gather}
\[
C_{i} \doteq \half \langle \, {\bf{y}}_i,{\bf{y}}_i \, \rangle.
\]
The integrals $\{H_{i}^r\}_{i=1}^{N}$ are first integrals of
motion and the integrals $\{C_{i}\}_{i=1}^{N}$ are the Casimir
functions given in equation~\eqref{poi}.
\end{proposition}

\begin{proposition} The curve $\det( \CL_{\CG}^{t}(\l) -\mu \, \mathds{1})=0$,
$\l,\mu \in \mathbb{C}$, with $\CL_{\CG}^t(\l)$ given in equation
\eqref{lgauretrig}, provides a  set of $2N$ independent involutive
integrals of motion given by
\begin{gather*}
H_{i}^t \doteq  \sum_{\stackrel{\scriptstyle{j=1}} {j \neq i}}^N
\frac{y_i^1 \, y_j^1 + y_i^2 \, y_j^2+ \cos(\l_i - \l_j)\, y_i^3
\, y_j^3}{\sin(\l_i - \l_j)},  \qquad \sum_{i=1}^N H_i^t =0,
\\
H_{0}^t \doteq \left(\sum_{i=1}^N y_i^3\right)^2, \qquad C_{i}
\doteq \half \langle \, {\bf{y}}_i,{\bf{y}}_i \, \rangle.
\end{gather*}
The integrals $\{H_{i}^t\}_{i=0}^{N}$ are first integrals of
motion and the integrals $\{C_{i}\}_{i=1}^{N}$ are the Casimir
functions given in equation \eqref{poi}.
\end{proposition}

\begin{proposition} The curve $\det( \CL_{\CG}^{e}(\l) -\mu \, \mathds{1})=0$,
$\l,\mu\in \mathbb{C}$, with $\CL_{\CG}^e(\l)$ given in equation
\eqref{lgaurell}, provides a  set of $2N$ independent involutive
integrals of motion given by
\begin{gather*}
 H_{i}^e \doteq  \sum_{\stackrel{\scriptstyle{j=1}} {j \neq i}}^N
\frac{{\rm dn} (\l_i -\l_j) \, y_i^1 \, y_j^1 + y_i^2 \, y_j^2+
{\rm cn} (\l_i -\l_j) \, y_i^3 \, y_j^3 }{{\rm sn} (\l_i - \l_j)},
\qquad \sum_{i=1}^N H_i^e =0,\nonumber \\
 H_{0}^e \doteq \sum_{\stackrel{\scriptstyle{i,j=1}} {i \neq j}}^N
\left[\,y_i^1 \, y_j^1 \,  g_1(\l_i -\l_j) + y_i^2 \, y_j^2 \,
g_2(\l_i -\l_j) + y_i^3 \, y_j^3 \, g_3(\l_i -\l_j) \, \right],
\nonumber
\end{gather*}
with
\[
g_1(\l) \doteq  \frac{\theta'_{11} \, \theta'_{10}(\l)}{
\theta_{10}\,
  \theta_{11} (\l)}, \qquad
g_2(\l) \doteq  \frac{\theta'_{11} \, \theta'_{00}(\l)}{
\theta_{00}
  \,\theta_{11} (\l)}, \qquad
g_3(\l) \doteq \frac{\theta'_{11} \, \theta'_{01}(\l)}{
\theta_{01}
  \,\theta_{11} (\l)},
\]
and
\[
C_{i} \doteq \half \langle \, {\bf{y}}_i,{\bf{y}}_i \, \rangle.
\]
Here $\theta_{\al \be} (\l)$, $\al,\be=0,1$, is the theta
function\footnote{We are using  the notation adopted in \cite{ST}:
\[
\theta_{\al \be} (\l) \doteq \theta_{\al \be} (\l,\tau) = \sum_{n
\in \mathbb{Z}} {\rm exp} \left[\,\pi \, \ri\left( n+
\frac{\al}{2}\right)^2 \tau + 2\, \pi \, \ri \left( n+
\frac{\al}{2} \right)\left( n+ \frac{\be}{2} \right) \, \right],
\] $\al,\be=0,1$, where $\tau$ is a complex number in the upper half plane.},
and $\theta_{\al \be} \doteq \theta_{\al \be}(0)$, $\theta'_{\al
\be} \doteq (d/d \l)_{\l=0} \theta_{\al \be}(\l)$. The integrals
$\{H_{i}^e\}_{i=0}^{N}$ are first integrals of motion and the
integrals $\{C_{i}\}_{i=1}^{N}$ are the Casimir functions given in
equation \eqref{poi}.
\end{proposition}

In the rational case it is possible to select a simple and
remarkable Hamiltonian. It is given by the following linear
combination of the integrals of motion $\{H_{i}^r\}_{i=1}^{N}$
given in equation (\ref{35}):
\begin{gather}
 \sum_{i=1}^N \eta_i \, H_i^r =\half
\sum_{\stackrel{\scriptstyle{i,j=1}} {i \neq j}}^N \frac{\eta_i -
\eta_j}{\l_i - \l_j} \, \langle \,  {\bf{y}}_i , {\bf{y}}_j \,
\rangle + \sum_{i=1}^N \eta_i \, \langle \,  {\bf{p}} , {\bf{y}}_i
\, \rangle,\label{itglrrrr}
\end{gather}
where the $\eta_i$'s with $\eta_i \neq \eta_k$, $1 \leq i,k \leq N$,
are arbitrary complex numbers. An interesting specialization of
the Hamiltonian (\ref{itglrrrr}) is obtained considering $\eta_i =
\l_i$, $1 \leq i \leq N$:
\begin{gather}
\CH_\CG^r \doteq \half \sum_{\stackrel{\scriptstyle{i,j=1}} {i
\neq j}}^N \langle \,  {\bf{y}}_i , {\bf{y}}_j \, \rangle +
\sum_{i=1}^N \l_i \, \langle \,  {\bf{p}} , {\bf{y}}_i \,
\rangle.\label{itglt}
\end{gather}

\begin{proposition} The equations of motion w.r.t.\ the Hamiltonian \eqref{itglt}
are given by
\begin{gather}
\dot {\bf y}_i = \left[\l_i\, {\bf{p}}  +  \sum_{j=1}^N
{\bf{y}}_j\,, \,{\bf{y}}_i \right], \qquad 1 \leq i \leq N,
\label{1g}
\end{gather}
where $\dot {\bf y}_i \doteq d {\bf y}_i  / dt$. Equations
\eqref{1g} admit the following Lax representation:
\begin{gather*}
\dot \CL_{\CG}^r (\l) =\big[\,  \CL_{\CG}^r(\l), \CM_{\CG}^{(r,-)}
(\l)  \, \big]=-\big[\,  \CL_{\CG}^r(\l), \CM_{\CG}^{(r,+)} (\l)
\, \big], \nonumber
\end{gather*}
with the matrix $\CL_{\CG}^r(\l)$ given in equation \eqref{lgaure}
and
\begin{gather}
\CM_{\CG}^{(r,-)} (\l) \doteq \sum_{i=1}^{N} \frac{\l_i \,
{\bf{y}}_i}{\l - \l_i}, \qquad \CM_{\CG}^{(r,+)} (\l) \doteq \l
\, {\bf{p}} + \sum_{i=1}^{N}{\bf{y}}_i . \label{100g}
\end{gather}
\end{proposition}

\begin{proof}
A direct computation.
\end{proof}

\section[Contraction of $su(2)$ Gaudin models: the two-body case]{Contraction of $\boldsymbol{\su2}$ Gaudin models: the two-body case}

In the present section we f\/ix $N=2$, namely we consider two-body
$\su2$ Gaudin models.

It is well-known that the In\"on\"u--Wigner contraction of $\su2
\oplus \su2$, i.e.\ a Lie algebra isomorphic to
$\mathfrak{so}(4)$, gives the real Euclidean algebra $\e3$
\cite{IW}. Let us def\/ine the isomorphism $\phi_\ep:
\mathfrak{su}^* (2) \oplus \mathfrak{su}^* (2) \rightarrow
\mathfrak{su}^* (2) \oplus \mathfrak{su}^* (2)$ by the map
\begin{gather}
\phi_\ep: ({\bf{y}}_1,{\bf{y}}_2 ) \longmapsto ({\bf{m}},{\bf{a}}
) \doteq ({\bf{y}}_1+ {\bf{y}}_2, \ep\, (\nu_1 \, {\bf{y}}_1+
\nu_2\, {\bf{y}}_2)), \label{r2r}
\end{gather}
where $\nu_1,\nu_2 \in \mathbb{C}$, $\nu_1 \neq \nu_2$ and $0 <
\ep \leq 1$ plays the role of a
 contraction parameter. In the limit $\ep \rightarrow 0$
the Lie--Poisson brackets on $\mathfrak{su}^* (2) \oplus
\mathfrak{su}^* (2)$ are mapped by $\phi_\ep$ into the
Lie--Poisson brackets on $\mathfrak{e}^* (3) \cong \mathfrak{su}^*
(2) \oplus_s \mathbb{R}^3$:
\begin{gather}
\big\{    m^{\alpha}, m^{\beta}   \big\}= -\epsilon_{\al \be
\gamma} \,m^{\gamma}, \qquad \big\{   m^{\alpha}, a^{\beta}
\big\}= -\epsilon_{\al \be \gamma} \, a^{\gamma}, \qquad \big\{
a^{\alpha}, a^{\beta}   \big\}= 0. \label{rr}
\end{gather}
Obviously, the map $\phi_\ep$ is not an isomorphism after the
contraction limit $\ep \rightarrow 0$. The Lie--Poisson brackets
(\ref{rr})  are degenerate: they possess the two Casimir functions
\begin{gather}
K_1\doteq  \langle\,  {\bf{m}}, {\bf{a}} \,\rangle, \qquad K_2
\doteq  \half \langle \,{\bf{a}}, {\bf{a}} \,\rangle. \label{casi}
\end{gather}
A direct calculation shows that if $H( {\bf{y}}_1, {\bf{y}}_2)$
and $G( {\bf{y}}_1, {\bf{y}}_2)$ are two involutive functions
w.r.t. the Lie--Poisson brackets on $\mathfrak{su}^* (2) \oplus
\mathfrak{su}^* (2)$ then, in the contraction limit $\ep
\rightarrow 0$, the functions $\phi_\ep(H( {\bf{y}}_1,
{\bf{y}}_2))$ and $\phi_\ep(G( {\bf{y}}_1, {\bf{y}}_2))$ are in
involution w.r.t.\ the Lie--Poisson brackets on $\mathfrak{e}^*
(3)$.

Our aim is now to apply the contraction map $\phi_\ep$ def\/ined
in equation (\ref{r2r}) to the Lax matrices of the two-body $\su2$
Gaudin models, i.e. the matrices in equations (\ref{lgaure}),
(\ref{lgauretrig}) and (\ref{lgaurell}) with $N=2$. To do this a
second ingredient is needed: as shown in \cite{KPR,MPR,MPRS2,P} we
have to consider the {\it pole coalescence} $\l_i = \ep \, \nu_i$,
$i=1,2$. This fusion procedure can be considered as the analytical
counterpart of the algebraic contraction given by the map in
equation (\ref{r2r}).

A straightforward computation leads to the following statement
\cite{KPR,P}.

\begin{proposition}
In the limit $\ep \rightarrow 0$, the isomorphism \eqref{r2r} maps
the Lax matrices \eqref{lgaure}, \eqref{lgauretrig}
and~\eqref{lgaurell} with $\l_i = \ep \, \nu_i$, $i=1,2$,
respectively into the Lax matrices
\begin{gather}
 \CL^r(\l)  \doteq  {\bf p}+ \frac{ {\bf m}}{\l} +\frac{ {\bf a}}{\l^2},
\label{lgaurerat} \\
 \CL^t(\l)  \doteq
\frac{1}{\sin (\l)} \left[ \, \sig_1\, m^1 + \sig_2\, m^2 + \cos(\l) \, \sig_3\, m^3  \, \right] \nonumber    \\
 \phantom{ \CL^t(\l)  \doteq}  {}+ \frac{1}{\sin^2 (\l)} \left[ \, \cos(\l) \, (\sig_1\, a^1 + \sig_2\, a^2) + \sig_3\, a^3  \, \right],  \label{lgauretrigy} \\
 \CL^e(\l)  \doteq
\frac{1}{{\rm sn}(\l)} \left[ \, {\rm dn}(\l) \, \sig_1\, m^1+ \sig_2\, m^2+{\rm cn}(\l)\, \sig_3\, m^3 \, \right] \nonumber  \\
 \phantom{\CL^e(\l)  \doteq}{}   + \frac{1}{{\rm sn}^2(\l)} \left[ \, {\rm cn}(\l) \, \sig_1\, a^1+
{\rm cn}(\l)\, {\rm dn}(\l)\,  \sig_2\, a^2+{\rm dn} (\l)\,
\sig_3\, a^3 \, \right] .\label{lgaurellll}
\end{gather}
\end{proposition}

The Lax matrices given in equations (\ref{lgaurerat}),
(\ref{lgauretrigy}) and (\ref{lgaurellll}) describe completely
integrable systems on the Lie--Poisson manifold associated with
$\mathfrak{e}^*(3)$. The remarkable feature of the above procedure
is that the contracted models inherit the linear $r$-matrix
algebra  (\ref{pb4}) of the ancestor system. The following
proposition holds \cite{KPR,MPR,MPRS2}.

\begin{proposition}
The Lax matrices $ \CL^r(\l)$, $\CL^t(\l)$, $\CL^e(\l)$ given in
equations \eqref{lgaurerat}, \eqref{lgauretrigy} and
\eqref{lgaurellll} satisfy the linear $r$-matrix algebra
\begin{gather}
\left\{ \CL^{r,t,e}(\la) \otimes \mathds{1}, \mathds{1} \otimes
\CL^{r,t,e}(\mu) \right\}+\left[\,r_{r,t,e}(\la-\mu),
\CL^{r,t,e}(\la) \otimes \mathds{1}+\mathds{1} \otimes
\CL^{r,t,e}(\mu)\, \right]=0, \label{PPO}
\end{gather}
for all $\l, \mu  \in \mathbb{C}$, with $r_{r,t,e}(\l)$ given in
equation \eqref{pb43ui}.
\end{proposition}

\subsection[A Lagrange top arising from the rational $su(2)$ Gaudin model]{A Lagrange top arising from the rational $\boldsymbol{\su2}$ Gaudin model}

Recall that the (3-dimensional) Lagrange case of the rigid body
motion around a f\/ixed point in a homogeneous f\/ield is
characterized by the following data: the inertia tensor is given
by ${\rm diag} (1,1,\alpha)$, $\al \in \mathbb{R}$, which means
that the body is rotationally symmetric with respect to the third
coordinate axis, and the f\/ixed point lies on the symmetry axis
\cite{A,BS,KPR,RSTS}.

As noticed in \cite{KPR} the Lagrange top can be obtained from the
two-body rational $\su2$ Gaudin model performing the contraction
procedure previously described.

Let us recall the main features of the dynamics of the Lagrange
top (in the rest frame). The equations of motion are given by:
\begin{gather}
\dot {{\bf{m}}} = [\,{\bf{p}},{\bf{a}} \, ], \qquad
\dot {{\bf{a}}} = [\,{\bf{m}} , {\bf{a}} \, ], \label{1000y}
\end{gather}
where ${\bf{m}}\in \mathbb{R}^3$ is the vector of kinetic momentum
of the body, ${\bf{a}}\in \mathbb{R}^3$ is the vector pointing
from the f\/ixed point to the center of mass of the body and
${\bf{p}}\doteq (0,0,p)$ is the constant vector along the external
f\/ield. An external observer is mainly interested in the motion
of the symmetry axis of the top on the surface $\langle \,
{\bf{a}}, {\bf{a}}\, \rangle=\mbox{constant}$. For an actual integration
of this f\/low in terms of elliptic functions see \cite{GZ}.

A remarkable feature of the equations of motion (\ref{1000y}) is
that they do not depend explicitly on the anisotropy parameter
$\al$ of the inertia tensor \cite{BS}. Moreover they are
Hamiltonian equations with respect to the Lie--Poisson brackets of
$\mathfrak{e}^*(3)$, see equation  (\ref{rr}). The Hamiltonian
function that generates the equations of motion (\ref{1000y}) is
given by
\begin{gather}
I_1^{r} \doteq \half \langle \, { \bf{m}}, {\bf{m}} \, \rangle +
\langle \,{\bf{p}}, {\bf{a}} \, \rangle, \label{Hlag}
\end{gather}
and the complete integrability of the model is ensured by the
second integral of motion $I_2^{r} \doteq \langle \,{\bf{p}},
{\bf{m}} \, \rangle$. These involutive Hamiltonians can be
obtained by computing the spectral invariants of the Lax matrix
given in equation (\ref{lgaurerat}). The remaining two spectral
invariants are given by the Casimir functions of the Lie--Poisson
brackets of $\mathfrak{e}^*(3)$, see equation (\ref{casi}).

\begin{proposition} The Hamiltonian flow \eqref{1000y} generated by the Hamiltonian \eqref{Hlag}
admits the following Lax representation:
\begin{gather}
\dot \CL^r (\l) =\big[\,  \CL^r(\l), \CM^{(r,-)} (\l)  \, \big]= -
\big[\,  \CL^r(\l), \CM^{(r,+)} (\l)  \, \big], \nonumber
\end{gather}
with the matrix $\CL^r(\l)$ given in equation \eqref{lgaurerat}
and
\begin{gather}
\CM^{(r,-)}(\l) \doteq \frac{{\bf{a}}}{\l}, \qquad \CM^{(r,+)}(\l)
\doteq \l \,{\bf{p}} + {\bf{m}} . \label{100g3}
\end{gather}
\end{proposition}

\begin{proof}
A direct verif\/ication.
\end{proof}

\begin{remark} Using the contraction map (\ref{r2r}) one can obtain equations (\ref{1000y}) directly from equations
(\ref{1g}) (with $N=2$):
\begin{gather}
 \dot {{\bf{m}}} = \dot {{\bf{y}}}_1 + \dot {{\bf{y}}}_2 = [ \, {{\bf{p}}}, \ep
\,( \nu_1 \,  {{\bf{y}}}_1 + \nu_2 \,  {{\bf{y}}}_2) \, ]=
[\,{\bf{p}},{\bf{a}} \, ], \nonumber \\
 \dot {{\bf{a}}} = \ep (\nu_1 \, \dot {{\bf{y}}}_1 + \nu_2 \, \dot {{\bf{y}}}_2
) = [ \, {{\bf{y}}}_1 + {{\bf{y}}}_1 , \ep \,( \nu_1 \,
{{\bf{y}}}_1 + \nu_2 \,  {{\bf{y}}}_2) \, ] + O(\ep^2)
\xrightarrow{\ep \rightarrow 0} [\,{\bf{m}},{\bf{a}} \, ].
\nonumber
\end{gather}
Performing the same procedure on the Hamiltonian $\CH_\CG^r
\doteq  \l_1 \, H_1^r + \l_2 \, H_2^r$ given in equation
(\ref{itglt}) (with $N=2$) and on the linear integral $H_1^r +
H_2^r =\langle \,   {\bf{p}} , {\bf{y}}_1+{\bf{y}}_2  \, \rangle$
we recover the integrals of motion of the Lagrange top. We have:
\[
\CH_\CG^r = \half \langle \,{\bf{y}}_1 + {\bf{y}}_2, {\bf{y}}_1+
{\bf{y}}_2 \, \rangle - C_1 -C_2 + \langle \, {\bf{p}}, \ep \,
\nu_1 \, {\bf{y}}_1 + \ep \, \nu_1 \, {\bf{y}}_2 \, \rangle ,
\]
being $C_1 \doteq \langle \, {\bf{y}}_1 ,  {\bf{y}}_1 \, \rangle
/2 $, $C_2\doteq \langle \, {\bf{y}}_2 ,  {\bf{y}}_2 \, \rangle /2
$ just Casimir functions. Hence,
\[
\CH_\CG^r  \xrightarrow{\ep \rightarrow 0} \half \langle \,
{\bf{m}},  {\bf{m}} \, \rangle + \langle \,{\bf{p}}, {\bf{a}} \,
\rangle = I_1^{r}.
\]
Finally, $ H_1^r + H_2^r = \langle \,   {\bf{p}} ,
{\bf{y}}_1+{\bf{y}}_2  \, \rangle =\langle \,   {\bf{p}} ,{\bf{m}}
\, \rangle = I_2^{r} $. The same procedure allows one to recover
the auxiliary matrices $\CM^{(r,\pm)}(\l)$ given in equation
(\ref{100g3}) from the matrices $\CM_\CG^{(r,\pm)}(\l)$ given in
equation (\ref{100g}).
\end{remark}

\subsection[A Clebsch system arising from the elliptic $su(2)$ Gaudin model]{A Clebsch system arising from the elliptic $\boldsymbol{\su2}$ Gaudin model}

Let us now consider the Lax matrix given in equation
(\ref{lgaurellll}) obtained performing the contraction procedure
on the Lax matrix of the $\su2$ elliptic Gaudin model with $N=2$.

A direct computation shows that the spectral invariants of
$\CL^e(\l)$ are given by the following quadratic functions:
\begin{gather}
 I_1^e  \doteq \half \langle\,  {\bf{m}}, {\bf{m}} \,\rangle -\half
\langle\,  {\bf{a}}, B_1 \, {\bf{a}} \,\rangle,
\label{te2} \\
 I_2^e \doteq \half \langle\,  {\bf{m}}, A \, {\bf{m}} \,\rangle -\half
\langle\,  {\bf{a}}, B_2 \, {\bf{a}} \,\rangle,
\label{te3}  \\
 K_1\doteq  \langle\,  {\bf{m}}, {\bf{a}} \,\rangle, \qquad
 K_2 \doteq  \half \langle \,{\bf{a}}, {\bf{a}} \,\rangle,\nonumber
\end{gather}
where
\begin{gather*}
 B_1 \doteq {\rm {diag}}(0,k^2,k^2-1), \qquad
 B_2 \doteq {\rm{diag}}(0,0,k^2-1),\qquad
 A \doteq {\rm {diag}}(1-k^2,1,0).
\end{gather*}
Obviously, the choice $k=0$ in the integrals (\ref{te2}) and
(\ref{te3}) provides the spectral invariants of the trigonometric
Lax matrix $\CL^t(\l)$ given in equation (\ref{lgauretrigy}). Thus
the system described by~$\CL^t(\l)$ is a~subcase of the one
described by~$\CL^e(\l)$. The quadratic functions (\ref{te2}) and
(\ref{te3}) are in involution w.r.t. the Lie--Poisson brackets on
$\mathfrak{e}^*(3)$ thanks to the $r$-matrix formulation in
equation (\ref{PPO}).

Let us now recall the main features of the (3-dimensional) Clebsch
case of the free rigid body motion (in an ideal f\/luid)
\cite{RSTS,Su}. This problem is traditionally described by a
Hamiltonian system on $\mathfrak{e}^*(3)$ with the Hamiltonian
function
\begin{gather}
H \doteq \half \langle\,  {\bf{m}}, \CA \, {\bf{m}} \,\rangle-
\half \langle\,  {\bf{a}}, \CB \, {\bf{a}} \,\rangle, \label{cle}
\end{gather}
where $({\bf{m}}, {\bf{a}}) \in \mathfrak{e}^*(3)$ and the
matrices $\CA \doteq {\rm {diag}}(\alpha_1,\alpha_2,\alpha_3)$ and
$\CB \doteq {\rm
 {diag}}(\beta_1,\beta_2,\beta_3)$ are such that the following relation holds:
\[
\frac{\beta_1 - \beta_2}{\alpha_3} + \frac{\beta_2 -
\beta_3}{\alpha_1} + \frac{\beta_3 - \beta_1}{\alpha_2}=0,
\]
namely
\begin{gather}
\alpha_1 =\frac{\beta_2 -\beta_3}{\gamma_2 -\gamma_3}, \qquad
\alpha_2 =\frac{\beta_3 -\beta_1}{\gamma_3 -\gamma_1}, \qquad
\alpha_3 =\frac{\beta_1 -\beta_2}{\gamma_1 -\gamma_2},
\label{condr}
\end{gather}
for some matrix $\mathcal{C} \doteq {\rm
  {diag}}(\gamma_1,\gamma_2,\gamma_3)$.

Taking into account equations (\ref{te2})--(\ref{te3}) and
(\ref{condr}) we see that  $\mathcal{C} = {\rm
  {diag}}(0,k^2,k^2-1) =B_1$ for the Hamiltonian (\ref{te2}) and
$\mathcal{C} = {\rm
  {diag}}(1-k^2,1,0)=A$ for the Hamiltonian (\ref{te3}). Thus
$\CL^e(\l)$ can be considered as the Lax matrix of a special case
of the Clebsch system described by the Hamilton function
(\ref{cle}).

We now derive Lax representations for the Hamiltonian f\/lows
corresponding to the Hamilton functions (\ref{te2})--(\ref{te3}).
They can be written in terms of $\su2$ matrices with an elliptic
dependence on the spectral parameter.

 The equations of motion w.r.t.\ the integrals $I_1^e$ and $I_2^e$ read respectively
\begin{gather}
\dot {{\bf{m}}} = \left[\,{\bf{a}}, B_1 \, {\bf{a}} \, \right], \qquad
\dot {{\bf{a}}} = \left[\,{\bf{m}} , {\bf{a}} \, \right],
\label{1000yell}
\end{gather}
and
\begin{gather}
\dot {{\bf{m}}} = \left[\,A \, {\bf{m}}, {\bf{m}} \, \right]
+\left[\,
  {\bf{a}}, B_2  \, {\bf{a}} \, \right] ,\qquad
\dot {{\bf{a}}} = \left[\,A \, {\bf{m}} , {\bf{a}} \, \right].
\label{1000yell2}
\end{gather}

A straightforward computation leads to the following result.

\begin{proposition} The Hamiltonian flow \eqref{1000yell} generated by the Hamiltonian \eqref{te2}
admits the Lax representation:
\begin{gather*}
\dot \CL^e (\l) =\left[\,  \CL^e(\l), \CM^e_1 (\l)  \, \right],
\end{gather*}
with the matrix $\CL^e(\l)$ given in equation \eqref{lgaurellll}
and
\begin{gather*}
\CM^e_1 (\l) \doteq \frac{1}{{\rm sn}(\l)} \left[ \, {\rm dn}(\l)
\, \sig_1\, a_1+ \sig_2\, a_2+{\rm cn}(\l)\, \sig_3\, a_3 \,
\right].
\end{gather*}

The Hamiltonian flow \eqref{1000yell2} generated by the
Hamiltonian \eqref{te3} admits the Lax representation:
\begin{gather*}
\dot \CL^e (\l) =\left[\,  \CL^e(\l), \CM^e_2 (\l)  \, \right],
\end{gather*}
with the matrix $\CL^e(\l)$ given in equation \eqref{lgaurellll}
and
\begin{gather*}
\CM^e_2 (\l) \doteq
\frac{1}{{\rm sn}^2(\l)} \left[ \, {\rm cn}(\l) \, \sig_1\, m_1+ {\rm cn}(\l)\, {\rm dn}(\l)\,\sig_2\, m_2+{\rm dn}(\l)\, \sig_3\, m_3 \, \right]   \\
 \phantom{\CM^e_2 (\l) \doteq}{}   + \frac{1}{{\rm sn}^3(\l)} \left\{ \, {\rm dn}(\l) \, \sig_1\, a_1+
{\rm dn}^2(\l)\,\,  \sig_2\, a_2+ {\rm cn}(\l) \,[{\rm dn}^2(\l) +
{\rm sn}^2(\l)] \, \sig_3\, a_3 \, \right\} .
\end{gather*}
\end{proposition}

\begin{remark}
We note that the ``traditional" Lax representations for the
Hamiltonian f\/lows (\ref{1000yell})--(\ref{1000yell2}) are given
in terms of Lax matrices depending rationally on the spectral
parameter \cite{RSTS,Su}. However a Lax representation with
elliptic dependence on the spectral parameter for the Clebsch
system is already known \cite{BB}. Hence the novelty of our
results consists just in establishing of the connection between
$\su2$ elliptic Gaudin models and the Clebsch system.

\end{remark}

\section{Integrable chains of interacting tops}

As shown in \cite{MPR,MPRS3,P} one can construct integrable
many-body systems starting with the one-body Lax matrices given in
equations (\ref{lgaurerat}), (\ref{lgauretrigy}) and
(\ref{lgaurellll}). Such systems describe completely integrable
(long-range) chains of interacting tops on the Lie--Poisson
manifold associated with $\oplus^M \mathfrak{e}^*(3)$, being $M$
the number of tops appearing in the chain. Moreover they admit the
same linear $r$-matrix formulation given in equation (\ref{pb4})
\cite{MPR,P}.

Let us denote with $({\bf{m}}_{i},{\bf{a}}_{i}) \doteq( m^1_{i},
m^2_{i}, m^3_{i},a^1_{i},a^2_{i},a^3_{i}) \in \mathfrak{e}^*(3)$
the pair of $\mathbb{R}^3$ vectors associated with the $i$-th top
of the chain. Thus the Lie--Poisson brackets on $\oplus^M
\mathfrak{e}^*(3)$ read
\begin{gather*}
\big\{ m^{\alpha}_{i}, m^{\beta}_{j}   \big\}=- \delta_{i,j} \,
\epsilon_{\al \be \gamma} \,m^{\gamma}_i, \qquad \big\{
m^{\alpha}_{i}, a^{\beta}_{j}    \big\}=- \delta_{i,j} \,
\epsilon_{\al \be \gamma} \,a^{\gamma}_i, \qquad
\big\{ a^{\alpha}_{i}, a^{\beta}_{j}    \big\}= 0, 
\end{gather*}
with $1 \leq i,j \leq M$. The above brackets are degenerate: they
possess the following $2M$ Casimir functions:
\begin{gather}
C_i^{(1)} \doteq  \langle \, {\bf{m}}_i, {\bf{a}}_i \, \rangle,
\qquad C_i^{(2)} \doteq  \half \langle \, {\bf{a}}_i, {\bf{a}}_i
\, \rangle, \qquad 1 \leq i \leq M. \label{reer43}
\end{gather}

According to equations (\ref{lgaurerat}), (\ref{lgauretrigy}) and
(\ref{lgaurellll}) we can consider the following Lax matrices
def\/ined on $\su2 [\,\l,\l^{-1}]$:
\begin{gather}
 \CL_{M}^r(\l) \doteq  {\bf{p}}
+ \sum_{i=1}^{M} \CL^r_i(\l-\mu_i), \label{RR} \\
 \CL_{M}^t(\l) \doteq
\sum_{i=1}^{M} \CL^t_i(\l-\mu_i), \label{TT} \\
  \CL_{M}^e(\l) \doteq
\sum_{i=1}^{M} \CL^e_i(\l-\mu_i), \label{EE}
\end{gather}
where  the $\mu_i$'s with $\mu_i \neq \mu_k$, $1 \leq i,k \leq M$,
are complex parameters of the models. The Lax matrix
$\CL_{M}^r(\l)$ describes a system of $M$ interacting Lagrange
tops, called {\it{Lagrange chain}} in \cite{MPR}, while the
matrices $\CL_{M}^t(\l),\CL_{M}^e(\l)$ govern the dynamics of $M$
interacting Clebsch systems. The latter models can be called
{\it{Clebsch chains}}.

The following proposition holds \cite{MPR,P}.

\begin{proposition}
The Lax matrices $ \CL^r_M(\l)$, $\CL^t_M(\l)$, $\CL^e_M(\l)$
given in equations \eqref{RR}, \eqref{TT} \linebreak
and~\eqref{EE} satisfy the linear $r$-matrix algebra
\begin{gather*}
\big\{ \CL^{r,t,e}_M(\la) \otimes \mathds{1}, \mathds{1} \otimes
\CL^{r,t,e}_M(\mu) \big\}+\big[\,r_{r,t,e}(\la-\mu),
\CL^{r,t,e}_M(\la) \otimes \mathds{1}+\mathds{1} \otimes
\CL^{r,t,e}_M(\mu)\, \big]=0,
\end{gather*}
for all $\l, \mu  \in \mathbb{C}$, with $r_{r,t,e}(\l)$ given in
equation \eqref{pb43ui}.
\end{proposition}

We now construct the spectral invariants of the Lagrange chain and
of the Clebsch chain with $k=0$.

\subsection{The Lagrange chain}

The complete set of integrals of the model can be obtained in the
usual way. In fact, a straightforward computation leads to the
following statement.

\begin{proposition}
The  hyperelliptic curve $\det( \CL_{M}^r(\l) -\mu \,
\mathds{1})=0$, $\l,\mu\in \mathbb{C}$, with $\CL_{M}^r(\l)$ given
in equation \eqref{RR} reads
\begin{gather*}
-\mu^2= \frac{1}{4} \langle \,{\bf{p}}, {\bf{p}} \, \rangle +
\half \sum_{i=1}^M \left[\, \frac{R_i^r}{\l
-\mu_i}+\frac{S_i^r}{(\l -\mu_i)^2}+ \frac{C_i^{(1)}}{(\l
-\mu_i)^3}+ \frac{C_i^{(2)}}{(\l -\mu_i)^4} \, \right],
\end{gather*}
where
\begin{gather*}
R_i^r  \doteq \langle \,{\bf{p}}, {\bf{m}}_i \, \rangle +
\sum_{\stackrel{\scriptstyle{j=1}} {j \neq i}}^M \left[\, \frac{
\langle \,{\bf{m}}_i, {\bf{m}}_j \, \rangle}{\mu_i - \mu_j}+
\frac{ \langle \,{\bf{m}}_i, {\bf{a}}_j \, \rangle  -  \langle
\,{\bf{m}}_j, {\bf{a}}_i \, \rangle } {(\mu_i - \mu_j)^2}
 - 2 \, \frac{ \langle \,{\bf{a}}_i, {\bf{a}}_j \, \rangle }{(\mu_i - \mu_j)^3}
\, \right],
\\
S_i^r  \doteq \langle \,{\bf{p}}, {\bf{a}}_i\, \rangle + \half
\langle \,{\bf{m}}_i, {\bf{m}}_i \, \rangle +
\sum_{\stackrel{\scriptstyle{j=1}} {j \neq i}}^M \left[\, \frac{
\langle \, {\bf{a}}_i, {\bf{m}}_j \, \rangle }{\mu_i - \mu_j}+
 \frac{ \langle \,{\bf{a}}_i, {\bf{a}}_j \, \rangle }{(\mu_i - \mu_j)^2}
\, \right].
\end{gather*}
The $2M$ independent integrals $\{R_{i}^r \}_{i=1}^{M}$ and
$\{S_{i}^r \}_{i=1}^{M}$ are involutive first integrals of motion
and the integrals $\{C_{i}^{(1)}\}_{i=1}^{M}$ and
$\{C_{i}^{(2)}\}_{i=1}^{M}$ are the Casimir functions given in
equation \eqref{reer43}.
\end{proposition}

Notice that, as in the $\su2$ rational Gaudin model, there is a
linear integral given by $ \sum\limits_{i=1}^M  R_i^r=
\sum\limits_{i=1}^M \langle \,{\bf{p}},{\bf{m}}_i \, \rangle. $ A
natural choice for a physical Hamiltonian describing the dynamics
of the model can be constructed considering a linear combination
of the Hamiltonians $\{R_{i} \}_{i=1}^{M}$ and $\{S_{i}
\}_{i=1}^{M}$ similar to the one considered for the rational
Gaudin model, see equation (\ref{itglrrrr}):
\begin{gather}
\CH_{M}^r \doteq \sum_{i=1}^M ( \mu_i \, R_i^r + S_i^r)=
 \sum_{i=1}^M \langle \,{\bf{p}}, \mu_i \, {\bf{m}}_i + {\bf{a}}_i\, \rangle+
\half \sum_{i,j=1}^M \langle \,{\bf{m}}_i, {\bf{m}}_j \, \rangle.
\label{H!}
\end{gather}
If $M=1$ the Hamiltonian (\ref{H!}) gives the sum of the two
integrals of motion of the Lagrange top. Our aim is now to f\/ind
the Hamiltonian f\/low generated by $\CH_{M}^r$ and its Lax
representation.

\begin{proposition} The equations of motion w.r.t.\ the Hamiltonian \eqref{H!}
are given by
\begin{gather}
\dot {\bf{m}}_i = \left[  {\bf{p}} , {\bf{a}}_i \, \right] +
\left[\, \mu_i \, {\bf{p}} + \sum_{j=1}^M {\bf{m}}_j , {\bf{m}}_i  \right]  , \nonumber\\
\dot {\bf{a}}_i = \left[\mu_i \, {\bf{p}} + \sum_{j =1}^M
{\bf{m}}_j
 , {\bf{a}}_i  \right]  ,
 \label{eq34}
\end{gather}
with $1 \leq i \leq M$. Equations \eqref{eq34} admit the following
Lax representation:
\begin{gather*}
\dot \CL_{M}^r (\l) =\left[\,  \CL_{M}^r(\l), \CM_{M}^{(r,-)} (\l)
\, \right]= -\left[\,  \CL_{M}^r(\l), \CM_{M}^{(r,+)} (\l)  \,
\right],
\end{gather*}
with the matrix $\CL_{M}^r(\l)$ given in equation \eqref{RR} and
\begin{gather*}
\CM_{M}^{(r,-)} (\l) \doteq \sum_{i=1}^{M} \frac{1}{\l-\mu_{i}}
\left[\, \mu_i \, {\bf{m}}_i + \frac{\l \, {\bf{a}}_i
}{\l-\mu_{i}} \, \right], \qquad \CM_{M}^{(r,+)} (\l) \doteq \l
\,{\bf{p}}  + \sum_{i=1}^{M}  {\bf{m}}_i  .
\end{gather*}
\end{proposition}

\begin{proof}
A direct computation.
\end{proof}

\subsection[The Clebsch chain: the case $k=0$]{The Clebsch chain: the case $\boldsymbol{k=0}$}

The complete set of integrals of motion of the Clebsch chain, with
$k=0$, is given in the following statement.

\begin{proposition}
The  curve $\det( \CL_{M}^t(\l) -\mu \, \mathds{1})=0$, $\l,\mu
\in \mathbb{C}$, with $\CL_{M}^t(\l)$ given in equation~\eqref{TT}
reads
\begin{gather*}
 -\mu^2= H^t_0 + \half \sum_{i=1}^M \big[
R_i^t \, \cot (\l -\mu_i) + S_i^t \cot^2 (\l -\mu_i)  \\
 \phantom{-\mu^2=}{} +  C_i^{(1)} \, \cot^3 (\l -\mu_i)+
C_i^{(2)} \, \cot^4 (\l -\mu_i)  \big],
\end{gather*}
where
\begin{gather*}
H_0^t   \doteq \half \sum_{i=1}^M \left[\, (m_i^1)^2 + (m_i^2)^2\,
\right]
- \half \sum_{\stackrel{\scriptstyle{i,j=1}} {i \neq j}}^M m_i^3 \, m_j^3 + \half \left(\sum_{i=1}^M a_i^3 \right)^2  \\
\phantom{H_0^t   \doteq}{} + \sum_{\stackrel{\scriptstyle{i,j=1}}
{i \neq j}}^M \frac{1}{\sin ( \mu_i -\mu_j)}
\left[\, a_i^1 \, m_j^1 +  a_i^2 \, m_j^2 + a_i^3 \, m_j^3 \, \cos ( \mu_i -\mu_j) \, \right] \\
\phantom{H_0^t   \doteq}{} + \half
\sum_{\stackrel{\scriptstyle{i,j=1}} {i \neq j}}^M \frac{\cot (
\mu_i -\mu_j)}{\sin ( \mu_i -\mu_j)}
\left[\, a_i^1 \, a_j^1 +  a_i^2 \, a_j^2 + a_i^3 \, a_j^3 \, \cos ( \mu_i -\mu_j) \, \right],\\
R_i^t  \doteq C_i^{(1)} + \half \sum_{\stackrel{\scriptstyle{j=1}}
{j \neq i}}^M \left(  m_i^3 \, a_j^3 -  m_j^3 \, a_i^3\right)
\\
\phantom{R_i^t  \doteq}{} + \sum_{\stackrel{\scriptstyle{j=1}} {j
\neq i}}^M \frac{1}{\sin ( \mu_i -\mu_j)} \left[\, m_i^1 \, m_j^1
+  m_i^2 \, m_j^2 + m_i^3 \, m_j^3 \, \cos ( \mu_i -\mu_j) \,
\right]
\\
\phantom{R_i^t  \doteq}{}+ \sum_{\stackrel{\scriptstyle{j=1}} {j
\neq i}}^M \frac{\cot ( \mu_i -\mu_j)}{\sin ( \mu_i -\mu_j)}
\big[\, m_i^1 \, a_j^1 +  m_i^2 \, a_j^2 + m_i^3 \, a_j^3 \, \cos
( \mu_i -\mu_j)
\\
\phantom{R_i^t  \doteq}{} - \, m_j^1 \, a_i^1 -  m_j^2 \, a_i^2 - m_j^3 \, a_i^3 \, \cos ( \mu_i -\mu_j)\, \big] \\
\phantom{R_i^t  \doteq}{} -2 \sum_{\stackrel{\scriptstyle{j=1}} {j
\neq i}}^M \frac{1}{\sin^3 ( \mu_i -\mu_j)} \left[\, a_i^1 \,
a_j^1 +  a_i^2 \, a_j^2 + a_i^3 \, a_j^3 \, \cos ( \mu_i -\mu_j)
\, \right],
\\
S_i^t  \doteq C_i^{(2)} +\half \left[\, (m_i^1)^2 + (m_i^2)^2 +
(m_i^3)^2 \, \right] + \half (a_i^3)^2 +
\half \sum_{\stackrel{\scriptstyle{i,j=1}} {j \neq i}}^M a^3_i \, a^3_j \\
\phantom{S_i^t  \doteq}{} + \sum_{\stackrel{\scriptstyle{j=1}} {j
\neq i}}^M \frac{1}{\sin ( \mu_i -\mu_j)}
\left[\, a_i^1 \, m_j^1 +  a_i^2 \, m_j^2 + a_i^3 \, m_j^3 \, \cos ( \mu_i -\mu_j) \, \right] \\
\phantom{S_i^t  \doteq}{} + \sum_{\stackrel{\scriptstyle{j=1}} {j
\neq i}}^M \frac{\cot ( \mu_i -\mu_j)}{\sin ( \mu_i -\mu_j)}
\left[\, a_i^1 \, a_j^1 +  a_i^2 \, a_j^2 + a_i^3 \, a_j^3 \, \cos
( \mu_i
  -\mu_j)\, \right].
\end{gather*}
The integrals $H_0^t$, $\{R_{i}^t \}_{i=1}^{M}$, $\{S_{i}^t
\}_{i=1}^{M}$ are involutive first integrals of motion (only $2M$
of them are independent). The integrals
$\{C_{i}^{(1)}\}_{i=1}^{M}$ and $\{C_{i}^{(2)}\}_{i=1}^{M}$ are
the Casimir functions given in equation~\eqref{reer43}.
\end{proposition}

\section{Concluding remarks and open problems}

In the present paper we have proposed an algebraic technique which
enabled us to derive two (3-dimensional) integrable cases of rigid
body dynamics (the Lagrange top and the Clebsch system) from
two-body $\su2$ Gaudin models. We remark that the explicit
construction of the Lagrange top starting from the $\su2$ rational
two-body Gaudin system has been presented for the f\/irst time in
\cite{KPR}. To the best of our knowledge the derivation of the
Clebsch system def\/ined by the involutive Hamiltonians
(\ref{te2})--(\ref{te3}) starting from the $\su2$ elliptic
two-body Gaudin system is new, although the novelty is essentially
in establishing of the connection between these two integrable
systems.

Let us stress that the construction outlined here is just a top of
an iceberg. In \cite{MPR,MPRS2,MPRS3,P} we presented a general and
systematic reduction, based on generalized In\"on\"u--Wigner
contractions, of classical Gaudin models associated with a simple
Lie algebra $\alg$. Suitable algebraic and pole coalescence
procedures performed on the $N$-pole Gaudin Lax matrices, enabled
us to construct one-body and many-body hierarchies of integrable
models sharing the same (linear) $r$-matrix structure of the
ancestor models. This technique can be applied to any simple Lie
algebra $\mathfrak{g}$ and whatever be the dependence (rational,
trigonometric, elliptic) on the spectral parameter. Fixing
$\mathfrak{g} = \mathfrak{su}(2)$, we constructed the so called
$\mathfrak{su}(2)$ {\it {hierarchies}} \cite{MPRS2,P}. In
particular the Lagrange top corresponds to the f\/irst element
($N=2$) of the $\mathfrak{su}(2)$ {\it {rational hierarchy}}, and
the Clebsch system is the f\/irst element of the
$\mathfrak{su}(2)$ {\it {elliptic hierarchy}}.

We studied also the problem of discretizing the Hamiltonian
f\/lows of the $\su2$ rational Gaudin model. One of the authors
(O.R.), together with Vadim Kuznetsov and Andy Hone, constructed
in \cite{HKR} one-point (complex) and two-point (real) B\"acklund
transformations (BTs) for this model. Later on, in \cite{KPR},
again in collaboration with Vadim, we studied the problem of
discretizing the dynamics of the Lagrange top using the BTs
approach \cite{KS1,KV}.

In \cite{P,PS}, using a dif\/ferent approach,  we have obtained a
new integrable discretization for the Hamiltonian f\/low given in
equation (\ref{1g}). It is expressed in terms of an explicit
Poisson map and a suitable contraction performed on it enables us
to construct discrete-time versions of the whole
$\mathfrak{su}(2)$ {\it {rational hierarchy}}. Our results
include, as a special case ($N=2$), the discrete-time version of
the Lagrange top found by Yu.B. Suris and A.I. Bobenko in
\cite{BS}. Moreover, the same procedure enabled us to f\/ind an
integrable discretization of the Hamiltonian f\/low (\ref{eq34}),
describing a discrete-time version of the Lagrange chain.

A natural extension of our discretizations could be the
construction of a suitable approach for models with a
trigonometric or elliptic dependence on the spectral parameter
instead of a~rational one. To the best of our knowledge there are
no results in this direction in literature. We remark here that
integrable discretizations for the f\/lows
(\ref{1000yell})--(\ref{1000yell2}) have been found by
Yu.B.~Suris, see \cite{Su,Su2,Su3}, by using rational Lax
matrices.

\subsection*{Acknowledgements}

M.P. thanks Yuri B. Suris and G. Satta for helpful comments. M.P.
was partially supported by the European Community through the FP6
Marie Curie RTN ENIGMA (Contract number MRTN-CT-2004-5652) and by
the European Science Foundation project MISGAM.

\newpage

\pdfbookmark[1]{References}{ref}
\LastPageEnding

\end{document}